\documentclass[12pt,preprint]{aastex}
\usepackage{emulateapj5}

\def\ecs{ergs cm$^{-2}$ s$^{-1} \, $}
\def\es{ergs s$^{-1} \ $ }

\def\lae{\mathrel{<\kern-1.0em\lower0.9ex\hbox{$\sim$}}}
\def\gae{\mathrel{>\kern-1.0em\lower0.9ex\hbox{$\sim$}}}

\def\1wga{\mbox{1WGA~J1216.9$+$3743}}
\def\fxv{\mbox{$f_X/f_V$}}
\def\cgs{\mbox{erg cm$^{-2}$ s$^{-1}$}}

\slugcomment{Version accepted by ApJ}

\shorttitle{1WGA~J1216.9$+$3743: a steep ULX in NGC~4244}

\shortauthors{Cagnoni et al.}

\begin{document}

\title{1WGA~J1216.9$+$3743: Chandra Finds an Extremely  Steep 
 Ultraluminous X-ray Source}
\author{I. Cagnoni\altaffilmark{1}, R. Turolla\altaffilmark{2}, A. Treves\altaffilmark{1}, J.-S. Huang\altaffilmark{3}, D.W. Kim\altaffilmark{3}, M. Elvis\altaffilmark{3} and A. Celotti\altaffilmark{4}}
\affil{$^{1}$ Dipartimento di Scienze, Universit\`a dell'Insubria,
Via Valleggio 11, Como, I-22100, Italy}
\affil{$^{2}$ Dipartimento di Fisica, Universit\`a di Padova, Via Marzolo 8,
I-35131, Padova, Italy}
\affil{$^{3}$ Harvard-Smithsonian Center for Astrophysics, 60 Garden Street,
Cambridge, MA 02138, USA}
\affil{$^{4}$ SISSA-ISAS, Via Beirut 2-4, I-34014, Trieste, Italy}
\email{ilaria.cagnoni@uninsubria.it}

\begin{abstract}
We report the discovery of  a new ultraluminous X-ray source (ULX)  
in the nearby galaxy NGC~4244  from
{\em Chandra} archival data. 
The source, \1wga,
 is one of the least luminous and softest ULXs discovered so far.
Its X-ray spectrum is the best available for a representative of
the soft ULXs, a class of sources 
recently discovered by {\em Chandra} and {\em XMM-Newton}.
 \1wga  appears  point-like in the {\em Chandra} image
and has a complex  spectral shape:
a multicolor disk model, suitable for brighter ULXs, is inadequate
for this source.
\1wga spectrum is heavily absorbed ($N_H \sim 1-4 \times 10^{21}$ cm$^{-2}$),
and very soft. The best-fit power-law model gives $\Gamma \sim 5$
 and implies a luminosity $L_{(0.5-10\, {\rm keV)}} \sim 10^{39}$ erg
s$^{-1}$.  
A comparison with previous detections shows that,
despite the  variability displayed by the source  during
the  {\em Chandra} observation,
\1wga \/ count rate, spectral shape and absorption are practically unchanged
over a 9-year period.
 
We performed also deep optical imaging of the field containing
the X-ray source, and found a possible ${\rm R}\sim 23.7$ counterpart. 
\end{abstract}


\section{Introduction}

X-ray observations of nearby galaxies have shown that their X-ray
emission comes from different classes of sources: X-ray
binaries, supernova remnants, hot interstellar medium in addition
to possible background active galactic nuclei (AGNs; see e.g. Fabbiano 1989
 for a review). Some of the
point-like sources appear to radiate well in excess of the Eddington
limit for a $1\, M_\odot$ object, with inferred luminosities in the
range $10^{39}-10^{40}$ \es. These objects  are often referred to as
ultra
luminous X-ray sources (ULXs) or intermediate-luminosity X-ray
objects (IXOs), inasmuch their luminosity is in between those of
``normal'' X-ray binaries and AGNs. Up to now about 90 ULXs have been
detected in more than 50 galaxies (see Colbert \& Ptak 2002 for a
recent catalog). ULXs seem to be preferentially located in the
outskirts of the host galaxy, although some are found in the
central regions (e.g. IXO 95 in NGC 6949). As pointed out by Colbert
\& Ptak, starburst galaxies contain a comparatively large number of
ULXs, but spirals seem not to be  favoured hosts with respect to
ellipticals. In a very recent paper, moreover, evidence was
found for a ULX associated with a globular cluster in NGC 4565 (Wu et
al. 2002). ULXs are often variable sources and their X-ray spectrum may be
quite soft (e.g. Makishima et al. 2000). No certain optical
counterpart has been found for these sources yet, implying an extreme
X-ray--to-optical flux ratio. Very recently Pakull
\& Mironi (2002)
reported the presence of emission nebulae at the position of some
ULXs and suggested that they could actually be related to the episode
which led to the ULX formation.

Thus far no generally accepted model has been presented to explain the huge
energy output of ULXs. Assuming that the Eddington limit is not
exceeded and emission is isotropic implies that ULXs are powered by
accretion onto a $\approx 50-100\, M_\odot$ black hole.
Present evolutionary
scenarios do not preclude such a possibility and the fact that these
are rare sources may be compatible with the mean number of ULX
per galaxy being only 1-2 (Colbert \& Ptak 2002). Alternatively ULXs
may be interpreted as conventional black hole binaries
($M_{BH}\lesssim 10\, M_\odot$) with modest beaming coming from a 
collimated X-ray emission ($b\sim 0.1$,
King et al. 2001) or to a relativistically beamed emission.
 In this picture the Galactic analogues of ULXs should be
the microquasars, such as  GRS 1915+105 and GRO 1655-40.

This paper presents a multi-wavelength study of a new ULX, \1wga,
found in NGC~4244. This source is one of the 16 peculiar {\em
ROSAT} PSPC sources selected for their extremely high
X-ray-to-optical flux ratio (Cagnoni et al. 2002). \1wga \/ is a  bright
($F_{(0.1-2.4\, {\rm keV})} > 10^{-13}$ \cgs )
 WGACAT (White, Giommi \& Angelini
1994) source  with blank fields, i.e. no optical
counterparts on the Palomar Observatory Sky Survey to O=21.5.
The extreme \fxv \/ ratio that follows is incompatible
with all major and common classes of extragalactic sources,
including normal quasars, AGNs,
normal galaxies and nearby clusters of galaxies (Maccacaro et
al. 1988).  Possibilities for the nature of these `blanks'
(Cagnoni et al. 2002)  include:
(a) Quasar-2s, i.e.  high luminosity, high redshift
heavily obscured quasars, the bright analogs of
Seyfert~2s;
(b) Low Mass Seyfert-2s, AGNs powered by a low-mass obscured
black hole (i.e. obscured Narrow Line Sy~1);
(c) AGNs with no big blue bump, e.g. low radiative efficiency flows;
(d) Isolated Neutron Stars (e.g. Treves et al. 2000);
(e) $\gamma$-ray burst X-ray afterglows or fast variable/ transient sources;
(f) failed clusters, in which a large overdensity of matter
has collapsed but has not formed galaxies  (Tucker, Tananbaum \& Remillard 1995);
(g) high redshift clusters of galaxies
and, most relevant to this paper,
(h) ULXs in nearby galaxies.

Using X-ray archival data ({\em ROSAT} and {\em Chandra}) and the information
obtained from optical and IR follow-ups,
we present strong evidence that \1wga \/ is indeed a ULX in
NGC~4244, a B=-18.4 (Olling 1996) edge-on spiral galaxy (Hubble type Scd)
at a distance of $\sim 3.6$~Mpc (Fry et al. 1999).

We present the  X-ray, optical and IR observations
in \S ~\ref{sec_observations} and discuss
the results in \S ~\ref{sec_discussion}.\\
Errors in the paper  represent 90\% confidence levels, unless
explicitly stated otherwise.

\section{Observations}
\label{sec_observations}

\subsection{{\em ROSAT}}

1WGA~J1216.9$+$3347 was serendipitously observed by {\em ROSAT} PSPC in
November 1991 and the results of this observation are reported in
Cagnoni et al. (2002); we summarize here the main findings.
During the 9~ks PSPC observation (600179n00) a total of 135 net
photons were collected between 0.07 and 2.4 keV.
We extracted a spectrum and fitted it with both an absorbed power-law and an
absorbed blackbody model.
Despite the scanty statistics and the large error bars both fits agree on the
extreme softness of \1wga \/ emission
($\Gamma =4.90^{+5.10}_{-1.74}$ for the power-law and $kT \sim 200$ eV
for the  blackbody models respectively) and on the presence of absorption
a factor of 10-30 in excess of the Galactic value in this direction
($N_{H{\rm Gal}} =1.69 \times 10^{20}$ cm$^{-2}$;
$N_H \sim 4.5$ and $\sim 1.5$ $\times 10^{21}$ cm$^{-2}$ for the power-law
and blackbody models, respectively).
The source absorbed flux in the $0.5-2.0$ keV band is
$\sim 1.2 \times 10^{-13}$ \cgs , which corresponds to an absorbed
luminosity of $\sim 2 \times 10^{38}$ erg~s$^{-1}$
and to an unabsorbed luminosity of $\sim 10^{39}$ erg~s$^{-1}$ for both the
 power-law and  black body models at NGC~4244 distance.
{\em ROSAT} PSPC lightcurve  is consistent with the source being constant during the observation (65\% probability).

\1wga \/ region was observed by the HRI instrument on board {\em ROSAT} in
June 1996 for $\sim 8.5$~ks, but the source was not detected,
consistently with the extrapolation in the HRI band of the best-fit
absorbed power-law. The $3 \sigma$ upper limit on the count rate in a
24$^{\prime \prime}$ circle is 0.0023 counts s$^{-1}$.

\subsection{{\em Chandra}}

1WGA~J1216.9$+$3347 was serendipitously observed by {\em Chandra} ACIS-S
on May 20 2000 during a $\sim 50$~ks observation of the nearby galaxy
NGC~4244.
\1wga \/ is 6~kpc off NGC~4244 center and
is detected as a point-like source on ACIS-S S2 front illuminated chip with a total of
1948 counts between 0.3 and 10 keV with peak position at
$\alpha = 12^{\rm h}\, 16^{\rm m}\, 56.927^{\rm s}$,
$\delta = +37^{\circ}43'35.89''$ (J2000).

We reprocessed {\em Chandra} archival data using the CXC analysis package
CIAO 2.2.1 (Elvis et al., in prep.) and CALDB 2.9. The source
position determination  was improved
by using the new geometry parameter file added to CALDB 2.9 and by removing the
pixel randomization of $\pm 0.5$ pixels introduced by the standard
{\em Chandra} pipeline processing (e.g. Garcia et al. 2001).
We corrected the data for bad pixels and checked for high particle
background times during the observation.
The observation is not affected by 
any significant background flare.
\1wga \/ average  count rate during this observation is
$(3.92 \pm 0.98) \times 10^{-2}$ photons s$^{-1}$ between 0.3 and 10 keV
for a total of 1948 counts in the $\sim 49$~ks of net exposure.
In a conservative test to confirm whether any background flare
might affect our results, we have  excluded the time intervals
with background count rate a factor 1.2 larger or smaller than the mean value
during the observation\footnote{Note that all the points lie within a factor 1.3}, as suggested by CIAO 2.2.1 Science Threads
 (http://asc.harvard.edu/ciao/threads/acisbackground/).
This would bring to an useful exposure time of 42~ks and to 1653 net
photons (0.3-10 keV) for \1wga . A detailed comparison shows that
 the count rate, spectrum and lightcurves 
are fully consistent with those extracted from the total 
observation\footnote{E.g. the 42~ks spectrum fitted with an absorbed power law model gives values perfectly in agreement with those reported in Table~1
$N_H = 4.46^{+0.51}_{-0.47} \times 10^{21}$ cm$^{-2}$, $\Gamma = 4.91^{+0.28}_{-0.25}$, normalization=$3.37^{+0.54}_{-0.44} \times 10^{-4}$ photons cm$^{-2}$ s$^{-1}$ keV$^{-1}$ and reduced $\chi ^2 =1.32$ for 39 degrees of freedom (d.o.f.). Errors represent here 1$\sigma$ confidence.}.

\begin{figure}
\epsscale{0.5}
\plotone{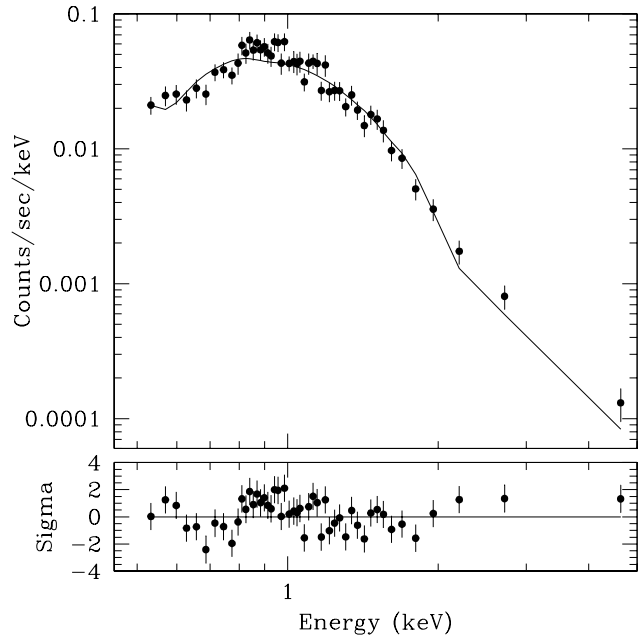}
\caption{\footnotesize{Spectrum and residuals to an absorbed power-law fit.
The residuals show that a more complex model is needed to model \1wga \/ 
emission.}\label{fig1}}
\end{figure}

We extracted a spectrum for the source using a circle with radius
$17.5^{\prime \prime}$ (35 pixels)  and a local background in an annular
region with radii of 20$^{\prime \prime}$  and 50$^{\prime \prime}$ (40 and 100
pixels) respectively from which a circle of 15$^{\prime \prime}$ centered
on a nearby source was excluded.
We binned the spectrum to have a minimum of
30 counts per bin and fitted it with SHERPA between 0.3 and 10.0 keV
with an absorbed power-law model and  an absorbed disk  blackbody.
Both models proved to be a poor representation of the data and
the results are unchanged when the 
ACIS quantum efficiency degradation is taken into account.
We tried with
more complex models, such as an absorbed power-law $+$ disk blackbody,
an absorbed power-law $+$ a Raymond-Smith thermal plasma and an absorbed
broken power-law and they are all acceptable.
The spectrum and the residuals to an absorbed power-law model are reported
in Figure~\ref{fig1} and  the results of all the fits are reported in
Table~1.
We also report in Table~1 the 0.5-10 keV absorbed fluxes and the 0.5-10 keV
absorbed and unabsorbed luminosities of the source.
It is clear from Table~1 that:
(i) \1wga \/ spectrum is extremely soft ($\Gamma \sim 4.90^{+0.42}_{-0.36}$);
(ii) simple models give an unacceptable fit to the data;
(iii) all but the broken power-law model predict a high column density
($\sim 1-7 \times 10^{21}$ cm$^{-2}$) in \1wga \/ direction;
(iv) \1wga \/ unabsorbed 0.5-10 keV luminosity is $\geq 10^{39}$ erg
s$^{-1}$ (for all but the broken power-law model, which could mimic the absorption with a flatter low energy slope and for the disk blackbody model, which 
is the statisticcally worst model).

\begin{figure}
\epsscale{0.5}
\plotone{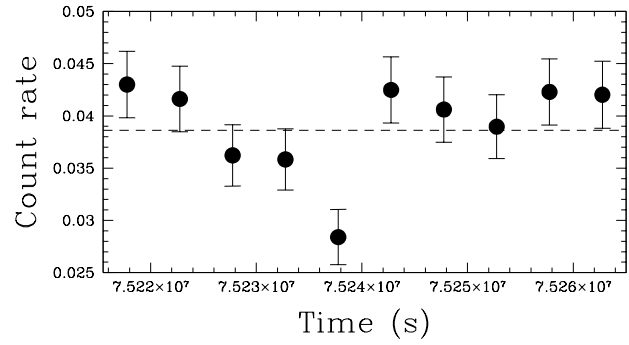}
\caption{\footnotesize{{\em Chandra} 0.3-10.0 keV 
lightcurve of \1wga \/ binned over 5000 s.
The dashed line represents the mean count rate during the observation
(i.e. $3.86 \times 10^{-2}$ counts s$^{-1}$)}\label{fig2}}
\end{figure}

We extracted  lightcurves for the source and background between 0.3 and
10 keV using the same circular and annular regions used for the spectral
analysis.
The background subtracted lightcurve of \1wga \/ binned over 5000~s is shown in
Figure~\ref{fig2} and the source displays a factor 1.5 variability.
A fit with a constant model gives a $\sim 0.5$\% probability for constant
emission, indicating that  \1wga  \/ is variable on timescales
$\leq$ 5000~s.
We have confirmed the variability by an independent method using a Bayesian-block analysis to the unbinned data which 
has recently been
developed for the {\em Chandra} Multi-wavelength Project (J. Drake, private 
communication; Kim et al. 2002, in preparation).

\subsection{Optical and Infrared}

We obtained deep  IR  imaging in the K-band (to $K\sim 18$)
with NSFCAM  at NASA/IRTF on January 2000 and deep R-band imaging
(to $R\sim 25$) with MOSAIC at KPNO 4m in February 2001.
We performed standard imaging reduction  using IRAF version 2.11.3 and
calculated  accurate astrometric solutions for the data using the 
digital sky survey.

The {\em Chandra} error box (0.67$^{\prime \prime}$ at 95\% confidence level
 for $\sim 1000$ net photons at $6^{\prime}$ off-axis, Kim et al. in preparation) contains only one object 
(at $0.6^{\prime \prime}$ from {\em Chandra} central position), 
which appears stellar on the R image (Figure~\ref{fig3}).
\1wga \/ counterpart lies in the optical extent of NGC~4244,
 at $\sim 6$ kpc from the galaxy center, and it is thus likely to be associated to the galaxy (we exclude a possible AGN in the background in
 \S ~\ref{sec_discussion}).

Since the night was not photometric we obtained an estimate of the source
magnitude rescaling from the USNO catalog sources in the 17.0--20.0
R-magnitude range falling on the CCD.
The estimated R-band magnitude is $\sim 23.7 \pm 0.5$, while the source is
not visible in the K-band image down to K$\sim 18$.
The X-ray over R-band flux ratio, defined as in Hornschemeir et al. (2001),
is $\sim 170$.

We searched for radio counterparts from the FIRST and the NVSS surveys but
none were found within $30^{\prime \prime}$ of
 the {\em Chandra} position, setting an upper limit of $\sim 1$ mJy at 1.4 GHz.

\section{Discussion}
\label{sec_discussion}

The properties of \1wga \/ indicate that it is
an ULX in  NGC~4244. The
source absorbed flux in the 0.5--2.0 keV band derived from
{\em Chandra} data ($\sim 1.8\times 10^{-13}$ \ecs) is consistent
within the errors with that measured with {\em ROSAT} ($\sim
1.2\times 10^{-13}$ \ecs), so the source appears not to have
significantly varied between the two observations ($\sim 9$ yrs).
The unabsorbed luminosity of \1wga \/ at NGC~4244
distance is $\sim 2\times 10^{39}$ \es for the absorbed power law model
Note however, that the 50~ks {\em Chandra} observation provides evidence 
for variability on a
much shorter timescale ($\sim 5000$ s). We also found that the
spectral shape did not change from the {\em ROSAT} observation and, although
no simple emission model gives a satisfactory fit, the
X-ray continuum appears to be very soft and can be roughly described by
a very steep absorbed power-law with $\Gamma\sim 5$ but not with thermal emission.
A complex spectral shape was also found for the stacked spectra obtained
from the ULXs in the Antennea galaxies by Zezas et al. (2002a)
(see their Figure~2): the spectrum presented here for \1wga \/ suggests that
the complexity might not be the effect of the superposition of different types of 
ULX spectra, but it is intrinsic to each source.

Using the X-ray flux at 1 keV from the absorbed power-law fit, the R-band magnitude and the  upper limit  at 1.4 GHz we computed \1wga \/
broad band spectral indices according to the formula
\begin{equation}
\alpha_{x_1 x_2} = - \frac{\log(f(x_2)/f(x_1))}{\log(\lambda(x_1)/
\lambda(x_2))} 
\end{equation}
We used the canonical 1~keV, V-band (5500 \AA ) and 5~GHz points, by extrapolating
the R-band flux density to the V-band and the 1.4 GHz flux density to 5
GHz assuming a flat spectral shape, and we obtain
$\alpha_{RO} \geq 0.60 $,  $\alpha_{OX} = 0.23$ and $\alpha_{RX} \geq 0.47$.
Such values place \1wga \/ out of the region occupied by AGNs
in the $\alpha_{RO}$--$\alpha_{OX}$ plane
(e.g. Caccianiga et al. 1999) since not even the most extreme BL Lac
objects
can reach such values (e.g. Costamante \& Ghisellini 2002).
Taking absorption into account would make the situation even more extreme
(e.g. an absorbing column of $\sim 4 \times 10^{21}$ cm$^{-2}$
 with Galactic dust to gas ratio would enhance the X-ray flux
by a factor of $\sim 10$ and the optical flux of a factor $\sim 6$).
We can thus exclude the  possibility of \1wga \/ being a background AGN.

We can also exclude the possibility of \1wga \/ being a supernova remnant
for the short timescale variability displayed
 in the {\em Chandra} observation and by the lack of a significant fading 
in the 9 years between the {\em ROSAT} and the {\em Chandra} observations.

Even if the best fit model for \1wga, a convex absorbed broken power-law,
is usually used to  describe the synchrotron peak observed in beamed sources
such as blazars, the sharp change in the power-law shape of \1wga \/  ($\Gamma _1 =0.22$, $\Gamma _2 = 3.85$) 
is difficult to reconcile with the smoothly curving blazar synchrotron peak 
(e.g. $\Gamma _1 = 2.1$, $\Gamma _2 = 2.8 $ 
measured for Mrk~421 by Guainazzi et al. 1999).

\1wga \/ luminosity falls at the lower end of the ULX range.
Some emission models implies $L_X < 10^{39}$ \es (see Table 1).
In particular, the absorbed power-law plus Raymond-Smith model,
and the absorbed broken power-law models,
which have the lowest $\chi^2$, give a luminosity of only
few $10^{38}$ \es. This would make \1wga \/ an ordinary
X-ray binary in NGC~4244. However, optically thin bremsstrahlung
emission is typical of extended sources, like clusters of galaxies,
and we found no evidence for a diffuse nature of this source in {\em
Chandra}
data and the broken power-law model does not have a straightforward
physical interpretation. The relatively low luminosity of \1wga  may
still suggest that this source is an X-ray
binary with a $M\approx 10\, M_\odot$ black hole, of the Cyg X-1 type.
The X-ray spectrum however strongly argues against this possibility, being much
softer of those of Galactic black hole candidates (BHCs).  BHCs spectra
in the high state are thermal and soft, but peak around a few keV and
extend up to $\sim 10$ keV.

The extreme softness of the X-ray spectrum
makes this source rather peculiar among the ULXs.
In fact, while very soft spectra have been already detected from other
ULXs with {\em ROSAT}, successive observations
over a wider energy range showed a  harder component.
An example of this behaviour is MS 0317.7$-$6647
 which has been first associated with
an isolated neutron star by Stocke et al. (1995) on the basis
of its soft, thermal spectrum. The source was later identified
with a ULX by Makishima et al. (2000) when {\em ASCA} data convincingly
showed a hard tail ($\Gamma \sim 2$), 
possibly associated with a multicolor disk
blackbody.
Before the advent of {\em Chandra} the ULXs seen by {\em ROSAT} and {\em ASCA} were
consistent with emission from a multicolor disk blackbody
(e.g. Makishima et al. 2000); {\em Chandra} is now discovering a new class of
steep ULX (e.g. out of the 30 ULX of Zezas et al. 2002b, 9 have a steep spectrum). \1wga \/ 
 is the source for which the best {\em Chandra} spectrum is available so far.
In the study of the Antennae galaxies Zezas et al. (2002a) find steeper spectra
for the least luminous ULX: \1wga seem to fit in this picture,
interpreted with the possible presence of undetected diffuse hot
interstellar medium in the proximity of the source.
Alternatively one could consider that  the indication of a luminosity 
dependent spectral shape is due 
to different accretion properties. 
 The steep ULXs  could be the ``galactic'' analog  of
Narrow Line Seyfert galaxies (a steep version of the Seyfert galaxies) 
which are 
thought to be powered by a lower mass black hole accreting at a higher
rate compared to ``normal'' Seyferts.

\begin{figure}
\epsscale{0.5}
\plotone{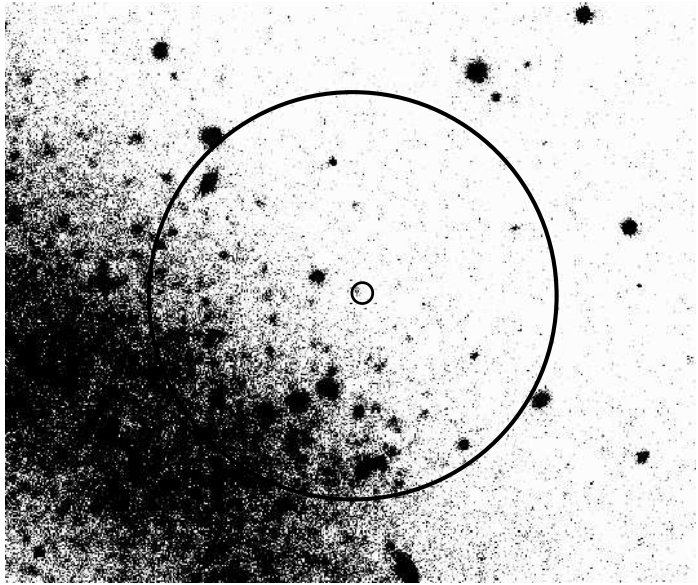}
\caption{\footnotesize{{\em ROSAT} PSPC ($r=39^{\prime \prime}$) and
 {\em Chandra}
 (enlarged to $r=2^{\prime \prime}$ for clarity) error circles on \1wga \/ positions. North is up and east is left in the image.}\label{fig3}}
\end{figure}

Follow-up optical observations allowed us to discover a possible counterpart
of \1wga \/ at $0.6^{\prime \prime}$ from {\em Chandra} position.
The association is based only on the positional coincidence.
The {\em Chandra} error box is quite small (95\% confidence radius 
$< 1''$, Kim et al. in preparation)
and the background field is not very crowded (see Fig.~\ref{fig3}), so the possibility of
a chance alignment appears unlikely, albeit it can not be ruled out on the
basis of present data. The optical counterpart has  $R\sim 23.7$;
if this source  is a star in NGC~4244, it has to be a red star or its bolometric correction
would imply a luminosity not compatible with even the most luminous 
stars known.
The most likely possibility is that the suggested counterpart is a late 
type giant/supergiant,
like a M5 II-III; for this spectral range the source
luminosity, corrected for extinction, would be $L\approx 7\times 10^{37}$
\es, about $10^4\, L_\odot$. The R-K$ < 5.7$ derived from the IR non-detection is not stringent and it
is satisfied  by the red stars.

ULXs are one of the most interesting class of sources which are now being 
investigated by the X-ray satellites {\em Chandra} and {\em XMM-Newton}.
We presented in this paper the best spectrum ($\sim 2000$ photons) 
available so far for a steep-spectrum ULX. 
Steep ULXs, like \1wga, ,
 appear to be related to the least luminous objects of the 
ULX class (e.g. Zezas et al. 2002a) and were essentially 
unknown before the launch of  {\em Chandra} and {\em XMM-Newton}.
Future investigations are needed to understand the emission mechanisms
powering these sources and to confirm and explain a possible luminosity 
 dependence  of the spectral shape.

\acknowledgments
This work made use of the Digitized Sky Surveys
produced at the Space Telescope Science Institute
under U.S. Government grant NAG W-2166 and of the 
US Naval Observatory catalog USNO-A1.0 manteined by the 
European Southern Observatory.
I.C. thanks Luigi Foschini and Andreas Zezas for the 
useful scientific discussion.
We are also grateful to the staffs of NASA-IRTF and FLWO. 
I.C. acknowledges a CNAA fellowship and A.C. the MIUR.

\newpage
\begin{rotate}
\begin{deluxetable}{l c c c c c c}
\footnotesize
\tablewidth{9.0in}
\tablecaption{Fit to the 0.3-10~keV {\em Chandra} spectrum with SHERPA}
\tablehead{
\colhead{Model}
&\colhead{$N_H$\tablenotemark{a}}
&\colhead{Parameters values}
&\colhead{$\chi ^2_{\nu}$/d.o.f.}
&\colhead{Abs.}
&\colhead{Abs.}
&\colhead{Unabs.}\\
\colhead{}
&\colhead{}
&\colhead{($E_b$, $kT$ in keV)}
&\colhead{}
&\colhead{$F_X$}\tablenotemark{b}
&\colhead{$L_X$}\tablenotemark{c}
&\colhead{$L_X$}\tablenotemark{c}\\
}
\startdata
Abs $\times$ power-law
&$4.19^{+0.75}_{-0.67}$
&$\Gamma = 4.90^{+0.42}_{-0.36}$  
$K=(3.38^{+0.82}_{-0.61})\times 10^{-4}$
&1.38/47
&2.31
&3.6
&21.6\\
\hline\\
Abs $\times$ disk blackbody
&$0.85^{+0.51}_{-0.47}$
&$kT = 0.300^{+0.030}_{-0.027}$  
$K=3.15^{+2.70}_{-1.42}$
&1.91/47
&2.19
&3.4
&4.6\\
\hline\\
Abs  $\times$ (power-law $+$ disk blackbody)
&$7.43^{+2.46}_{-1.51}$
&$\Gamma = 6.88^{+1.68}_{-0.71}$  
$K_{{\rm pow}}=(6.91^{+5.92}_{-2.91})\times 10^{-4}$
&1.28/45
&2.11
&3.3
&104.5\\
 &  &$kT = 0.712 \pm $  $K_{{\rm diskbb}}= 0.016^{+0.087}_{-0.015}$ & & & &\\
\hline\\
%
%
%
%
Abs  $\times$  broken power-law
&0.169\tablenotemark{d}
&$\Gamma _1 =0.22^{+0.46}_{-0.53}$  $\Gamma _2 =3.85^{+0.20}_{-0.18}$
&0.89/46
&2.15
&3.3
&3.5\\
& &$E_{b} =0.94 \pm 0.04$ $K=(1.71^{+0.29}_{-0.22}) \times 10^{-4}$ & & & &\\
\hline\\
Abs $\times$ (power-law +RS)
&$2.91^{+1.25}_{-0.96}$
&$\Gamma =4.26^{+0.86}_{-0.56}$  $K_{{\rm pow}} = 1.36^{+0.81}_{-0.48}\times 10^{-4}$
&1.05/45
&2.14
&3.3
&9.4\\
& & $kT=0.835^{+0.052}_{-0.079}$
 ${\rm Abun}= 0.3$\tablenotemark{e}  $K_{RS}=(1.53^{+0.56}_{-0.44}) \times 10^{-4}$ &&&&\\

\enddata
\label{tab_specfits}
\tablenotetext{a}{in units of $10^{21}$ cm$^{-2}$}
\tablenotetext{b}{Between 0.5 and 10 keV in units of $10^{38}$ erg cm$^{-2}$ s$^{-1}$}
\tablenotetext{c}{Between 0.5 and 10 keV in units of $10^{38}$ erg s$^{-1}$}
\tablenotetext{d}{The absorption was fixed to the Galactic value in \1wga \/
direction; if left as a free parameter it tends to zero}
\tablenotetext{e}{Similar values are obtained for Raymond-Smith abundance
equal to the solar value}
\end{deluxetable}
\end{rotate}

\end{document}